\documentclass[aip,jcp,preprint]{revtex4-1}
\usepackage{multirow}
\usepackage{epsfig}
\usepackage{amsfonts}
\usepackage{color}

\begin{document}

\title{The Plastic and Liquid Phases of CCl$_3$Br Studied by Molecular Dynamics Simulations}
\author{Nirvana B. Caballero }
\email{ncaballe@famaf.unc.edu.ar  }
\affiliation{Facultad de Matem\'atica, Astronom\'{\i}a y F\'{\i}sica,
Universidad Nacional de C\'ordoba, C\'ordoba, Argentina and IFEG-CONICET,
Ciudad Universitaria,
X5016LAE C\'ordoba, Argentina}
\author{Mariano Zuriaga }
\email{zuriaga@famaf.unc.edu.ar}
\affiliation{Facultad de Matem\'atica, Astronom\'{\i}a y F\'{\i}sica,
Universidad Nacional de C\'ordoba, C\'ordoba, Argentina and IFEG-CONICET,
Ciudad Universitaria,
X5016LAE C\'ordoba, Argentina}
\author{Marcelo Carignano}
\email{cari@northwestern.edu}
\affiliation{
Department of Biomedical Engineering and Chemistry of Life Processes Institute, Northwestern University,
 2145 Sheridan Road, Evanston, IL 60208, USA}
\author{Pablo Serra}
\email{serra@famaf.unc.edu.ar}
\affiliation{Facultad de Matem\'atica, Astronom\'{\i}a y F\'{\i}sica,
Universidad Nacional de C\'ordoba, C\'ordoba, Argentina and IFEG-CONICET,
Ciudad Universitaria,
X5016LAE C\'ordoba, Argentina}

\begin{abstract}
We present a molecular dynamics study of the liquid and plastic crystalline phases of CCl$_3$Br. We investigated the short-range orientational order using a recently developed classification method and we found that both phases behave in a very similar way. The only differences occur at very short molecular separations, which are shown to be very rare. The rotational dynamics was explored using time correlation functions of the molecular bonds. We found that the relaxation dynamics corresponds to an isotropic diffusive mode for the liquid phase, but departs from this behavior as the temperature is decreased and the system transitions into the plastic phase.
\end{abstract}
\date{\today}

\maketitle
\section{Introduction }

The plastic crystalline phase is a thermodynamic state characterized by long range translational order and orientational disorder (OD).\cite{parsonage78,sherwood79}  Nearly spherical molecules, like halogenomethane compounds of the form CCl$_n$Br$_{4-n}$, with $n$=$0,\dots,4$, are typical examples of plastic crystals. The formation of the OD phase is related to the little hindrance to the reorientational processes. This series of compounds offer the possibility of a systematic investigation of the effect of molecular symmetry, size and intermolecular interactions on the phase sequence and reorientational dynamics.

Tetrahalomethanes with $T_d$ molecular symmetry have been the subject of many experimental and computational studies\cite{gillen72,anderson79,cohen79,temleitner10,llanta01,pardoPRB2007,reyjcp2009,rey08a,rey08b,zuriaga09,zuriaga11,mcdonald82,dove86,tironi96,pettitt79}. Compounds with lower symmetry ($C_{3v}$ or $C_{2v}$) have attracted so far some attention from the experimental groups\cite{ohta95,binbrek99,parat05,zuriaga09}, but very little work has been done from a computational point of view.\cite{pothoczki10,pothoczki11} The polymorphism of CCl$_3$Br ($C_{3v}$ molecular symmetry) has been studied by several methods such as calorimetry\cite{ohta95} and neutron scattering,\cite{binbrek99} X-ray diffraction and densitometry\cite{parat05}. The low-temperature phase III (monoclinic) transforms at 238 K to the OD phase II (rhombohedral). On further heating, it transforms at 259 K to a {\em fcc} OD phase I. The {\em fcc} OD phase can be supercooled down to 230 K \cite{parat05}, and melts at 267 K.

In this paper we study CCl$_3$Br in its {\em fcc} plastic and liquid phases. Using molecular dynamics simulations we investigate the orientational structure and molecular relaxation characteristics in both phases. Our findings indicate a similar short-range orientational order in the two phases, with small differences occurring at very short distances. The relaxation dynamics, which appears to correspond to a random diffusive mode for the liquid phase, departs from this behavior for the plastic crystalline phase. 
The paper is organized as follows: Section II contains a brief discussion on the classification of the relative orientation of the molecules. Section III describes the model and the simulation methodology. The results and discussion about orientational order and rotational dynamics  are presented in section IV. Finally, Section VI contains our concluding remarks.

\section{Orientational Classes and Subclasses }

In a recent paper, Rey\cite{rey07} proposed a geometrical criterion to classify the relative orientation of two tetrahedral molecules (XY$_4$) in six well defined classes.  The classes are defined by considering two parallel planes, each one including one X atom from its corresponding molecule and both perpendicular to the line joining the two X atoms. Namely, for the case of CCl$_4$, the two parallel planes contain a C atom, and they are normal to the line defined by these two atoms. The six different classes are denoted by (1,1); (1,2); (1,3); (2,2); (2,3) and (3,3), where the numbers refer to the number of atoms from each molecule located between the two parallel planes. Following the original definition, we will call {\em corner}, {\em edge} and {\em face} to those configurations involving molecules contributing with one, two or three atoms to the region between the planes, respectively. Then, a (2,3) configuration is also referred to as an {\em edge-to-face} configuration. The classification can be extended to molecules of the type XY$_3$Z by defining subclasses. We characterize a subclass by the number ($k$) of Z atoms between the planes. Since there is only one Z atom per molecule, $k$ takes the values 0, 1 or 2 depending on the relative orientation of the molecules. Note that two molecules may contribute Z atoms adding up to $k$. For the case $k$=1 there is only one possible configuration of the class ($i$,$i$), and two possible configurations of the class ($i$,$j$). Then each one of the original Rey's classes is subdivided in three or four subclasses\cite{pothoczki11} explicitly defined in Table \ref{table_class} and denoted as ($i_m$,$j_n$), with $m+n=k$.

The probabilities for the six classes of molecules of the type  XY$_4$, calculated by Rey \cite{rey07} invoking the geometrical properties of the molecules and numerical integration, are shown in Table \ref{table_pro} below. The probabilities of occurrence of the different subclasses can be analytically calculated assuming that the relative orientations of the molecules are independent, or uncorrelated. This is the case if the spacial distance $r$ between the molecules is large; or strictly speaking, in the asymptotic limit for $r \rightarrow \infty$. The probability of each subclass is calculated by multiplying the probability of the corresponding class by the ratio between the number of configurations of the subclass to the total number of configurations of the class. The asymptotic values for the probabilities of the 21 subclasses are given in
Table \ref{table_pro}.

\begin{table*}[th!]
\begin{tabular}{|c|c|c|c|c|c|c|c}\hline\hline
\hspace{.1cm}  Subclass:  \hspace{.1cm}   &
\hspace{.2cm}  $(1_m,1_n)$  \hspace{.2cm}   &
\hspace{.2cm}  $(1_m,2_n)$  \hspace{.2cm}   &
\hspace{.2cm}  $(1_m,3_n)$  \hspace{.2cm}   &
\hspace{.2cm}  $(2_m,2_n)$  \hspace{.2cm}   &
\hspace{.2cm}  $(2_m,3_n)$  \hspace{.2cm}   &
\hspace{.2cm}  $(3_m,3_n)$  \hspace{.2cm}   \\ \hline \hline
$m=n=0$        & Y$-$Y          & Y$-$Y$_2$   &  Y$-$Y$_3$   &  Y$_2$$-$Y$_2$  &   Y$_2$$-$Y$_3$   & Y$_3$$-$Y$_3$   \\
\hline
$ m=0,n=1 $        & Y$-$Z          & Y$-$YZ          & Y$-$Y$_2$Z  & Y$_2$$-$YZ          &  Y$_2$$-$Y$_2$Z & Y$_3$$-$Y$_2$Z \\
$ m=1,n=0 $     &                      & Z$-$Y$_2$    & Z$-$Y$_3$    &                                  &  YZ$-$Y$_3$          &                                   \\
\hline
$m=n=1$        & Z$-$Z          & Z$-$YZ           & Z$-$Y$_2$Z  & YZ$-$YZ                 &  YZ$-$Y$_2$Z       & Y$_2$Z$-$Y$_2$Z  \\
\hline \hline
\end{tabular}
\caption{\label{table_class}
The configurations  of the 21 subclasses for XY$_3$Z molecules, where $m+n=k$
is the number of atoms of type Z between planes.
}
\end{table*}


\begin{table*}[!hb]
\begin{tabular}{|r|c|c|c|c|c|c|c}\hline\hline
           Class:       &  (1,1)       &    (1,2)    &      (1,3)       &   (2,2)       &        (2,3)   &       (3,3)     \\ \hline
  $P_c \simeq $  & 0.031 &  0.23  & 0.062     &  0.42    &  0.23     &  0.031   \\ \hline
\hline
 Subclass: &  $(1_m,1_n)$       &   $(1_m,2_n)$    &  $(1_m,3_n)$  &
 $(2_m,2_n)$    &  $(2_m,3_n)$    &   $(3_m,3_n)$  \\ \hline
$k=0$ &$\frac{9}{16} P_c\simeq0.017$  &   $ \frac{9}{24} P_c
\simeq 0.085$ & $\frac{3}{16} P_c\simeq0.012$&
$\frac{9}{36} P_c\simeq0.105$& $ \frac{3}{24} P_c\simeq0.028$ &
$\frac{1}{16} P_c\simeq0.0019$ \\    \hline
$k=1$ & $\frac{6}{16} P_c\simeq0.012$  &  $ \begin{array}{c}
\frac{9}{24} P_c\simeq 0.085\\  \frac{3}{24} P_c\simeq0.028
\end{array} $  & $ \begin{array}{c} \frac{9}{16} P_c\simeq0.035\\
\frac{1}{16} P_c\simeq0.0038\end{array}$  &
$ \frac{18}{36} P_c\simeq0.211$&$ \begin{array}{c} \frac{9}{24} P_c\simeq0.085\\ 
 \frac{3}{24} P_c \simeq0.028\end{array}$&  $\frac{6}{16} P_c\simeq0.012$
\\ \hline
$k=2$ & $\frac{1}{16} P_c\simeq0.0019$ &  $
\frac{3}{24} P_c\simeq0.028 $ & $ \frac{3}{16} P_c\simeq0.012$
& $ \frac{9}{36} P_c\simeq 0.115$& $  \frac{9}{24} P_c \simeq0.085$&
 $\frac{9}{16} P_c\simeq0.017$ \\   \hline
  \hline
\end{tabular}
\caption{\label{table_pro}  Asymptotic probabilities for the 6 classes  \cite{rey07} and the 21 subclasses of
configurations for XY$_3$Z molecules, where $k=m+n$ is the number of atoms of type Z  between planes.
The values correspond to configurations listed in Table \ref{table_class}.
}
\end{table*}



\section{Models and Methods}
The subject of our study corresponds to X=C, Y=Cl and Z=Br. We have modeled the CCl$_3$Br molecules as rigid tetrahedrons with the C atom at  the center, three Cl atoms on three vertices and a Br atom located in the remaining vertex. These are not regular tetrahedrons since the C-Cl and C-Br bond lengths differ. The intermolecular interactions were described by a Lennard-Jones potential plus a Coulombic term. The Lennard-Jones parameter for interactions between different atoms were calculated using the combination rule \mbox{$\epsilon_{ij}\,=\,\left(\,\epsilon_i \;\epsilon_{j} \right)^{1/2}$} and \mbox{$\sigma_{ij}\,=\,\frac{1}{2}\;\left(\,\sigma_i \,+\,\sigma_{j} \right)$}, where $i,j$ represent the three kinds of atoms C, Cl and Br. 
All interaction parameters and bond distances are summarized in Table \ref{tparameters}. Starting with a set of values taken from different sources, these parameters are the result of a refinement process: The charges of the C and Cl atoms are the result of {\em ab initio} calculations for CCl$_4$ \cite{rey00,llanta01} while the charge of the Br atom was chosen in order to reproduce the experimental value of the electric dipole moment of CCl$_3$Br, $\mu=0.21D$ \cite{Yama98,Miller56}. 
The Lennard-Jones parameters for C, and an initial guess for Cl and Br were obtained from the literature \cite{rey00,mcdonald82,evans85,barclay92}. A fine tuning of the L-J parameters for Cl and Br was done in a series of simulations trials so that the model system reproduces the experimental density at 273 K and 300 K \cite{parat05}. The initial configuration of the {\em fcc} plastic phase was prepared using a lattice constant $a=0.852$ nm.

We performed molecular dynamic (MD) simulations for a system of $N=4000$  molecules (20000 atoms) using the GROMACS v4.5.4 simulation package \cite{gromacs,gromacs4}.
The classical Newton's equations were integrated using the leap-frog algorithm with a time step of 0.001 ps.
The equilibration runs were done under $NPT$ conditions during 2 ns, and the pressure was controlled by a Parrinello-Rahman barostat with a time constant of 0.5 ps and a compressibility of $4.5 10^{-7}$ bar$^{-1}$. With the value obtained for the density in these runs we performed  2 ns production runs under $NVT$ conditions. 
Since the first non-zero electric interaction is a dipole-dipole force, we did not use long term correction for electrostatic  interactions. To validate this approach, we perform a few test runs including the time-expensive  Ewald summation
algorithm for electric charges in a smaller systems (500 molecules) with no appreciable difference in the results.

\begin{table}[!h]
\begin{tabular}{|c|c|c|c|}\hline\hline
\hspace{.5cm}  \mbox{} \hspace{.5cm} &  \hspace{.2cm} $\epsilon$ (kJ/mol) \hspace{.7cm}  &
\hspace{.5cm} $\sigma$ (nm) \hspace{.5cm} \mbox{} & \hspace{.5cm} $q$ (e) \hspace{.5cm} \\ \hline
C  &   $ 0.2276$ & 0.37739&  $-0.687$ \\ \hline
Cl  &  $ 1.40$  & $0.356$  &  0.170 \\ \hline
Br & $2.13 $ & $0.372 $&$ 0.177$ 
 \\ \hline\hline
$d$ (nm) &C$-$Cl: 0.1766 & C$-$Br: 0.1944 & Cl$-$Cl: 0.2884  \\
\hline \hline
\end{tabular}
\caption{\label{tparameters}  Model parameters for CCl$_3$Br.
}
\end{table}


\section{results and discussion}

\begin{figure}[ht]
\begin{center}
\includegraphics[width=0.4\textwidth]{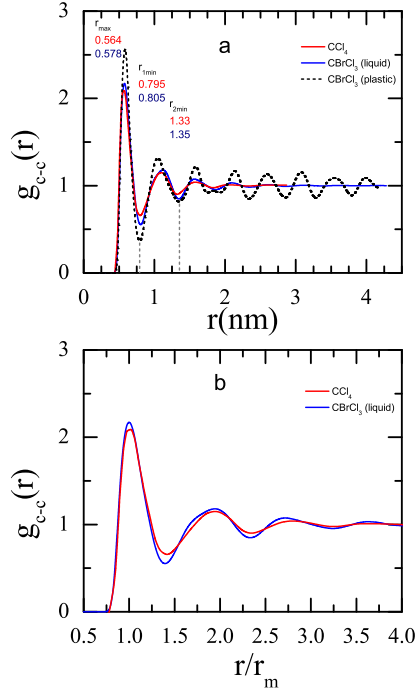}
\end{center}
\caption{  \label{gcc} (Color online) a) Calculated C$-$C radial distribution functions for CCl$_3$Br plastic crystal phase $T=220 K$ (dashed black line) and liquid phase $T=300 K$ (solid blue line); 
and CCl$_4$ from reference \cite{rey07} (solid red line). b) $g_{cc}(r)$ plotted as a function of the scaled distance (r/$r_m$) for the liquid phase.
} 
\end{figure}


\begin{figure}[bt]
\begin{center}
\includegraphics[width=0.4\textwidth]{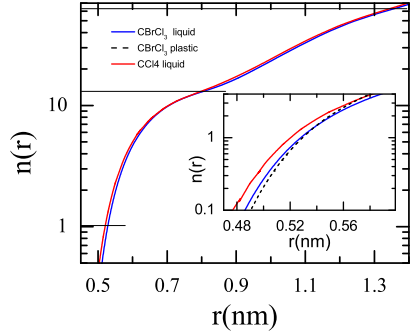}
\end{center}
\caption{  \label{mcn} (Color online) Molecular coordination numbers for CCl$_3$Br in liquid phase $T=300$ (solid blue line) and CCl$_4$ from reference \cite{rey07} (solid red line). The horizontal lines indicate the position of the first maximum, and first and second minima of the radial distribution function. Inset: molecular coordination number for distances where the local density of CCl$_3$Br in liquid phase (solid blue line) is higher than of the plastic phase (dashed black line).} 
\end{figure}


\subsection{Radial Distribution Functions}

In Figure \ref{gcc}a we show the carbon-carbon radial distribution functions for liquid and plastic phase of CCl$_3$Br. Since experimental results for CCl$_3$Br are unavailable, we include the radial distribution function corresponding to liquid  CCl$_4$ from reference \cite{rey07}, which is in excellent agreement with the findings of neutron scattering experiments. The $g(r)$ for the two molecular liquids show a similar behavior, with oscillations discernible up to $r \approx 2.5$ nm. The solid phase shows the typical behavior of the {\em  fcc} structure, and the long range correlations are reflected by the oscillating pattern that survives up to the largest distances. The only distinction between the structure of the two liquids are the small quantitative differences in the position of the maxima and minima in the radial distribution functions, which are indicated in Figure \ref{gcc}a. These differences reflect the different size of the Br and Cl atoms, making CCl$_3$Br effectively larger than CCl$_4$. In Figure \ref{gcc}b, we show the radial distribution functions for the two liquids, as a function of a the radial distance normalized by the position of the first peak. A common positional order emerges from this plot. The first minimum appears at $r = 1.4 \,r_m$, as observed in several other liquid of regular tetrahedral molecules\cite{reyjcp2009}.

In Figure \ref{mcn} we show the cumulative radial distribution functions for the same systems of Figure \ref{gcc}. The first and second coordination numbers are approximately 13 and 61 for the two cases. The insert of Figure \ref{mcn} shows details for short distances, including CCl$_3$Br in the {\em fcc} phase. As it was previously observed for CCl$_4$\cite{pardoPRB2007}, for short distances the radial distribution function for the liquid is larger that the one corresponding to the crystal. This is simply due to the greater thermal energy of the liquid that allows a closer approach between the molecules. This effect is very small, and in fact the two curves intersect before reaching their maxima.


\subsection{ Short Range Order}

\begin{figure}[bt]
\begin{center}
\includegraphics[width=0.4\textwidth]{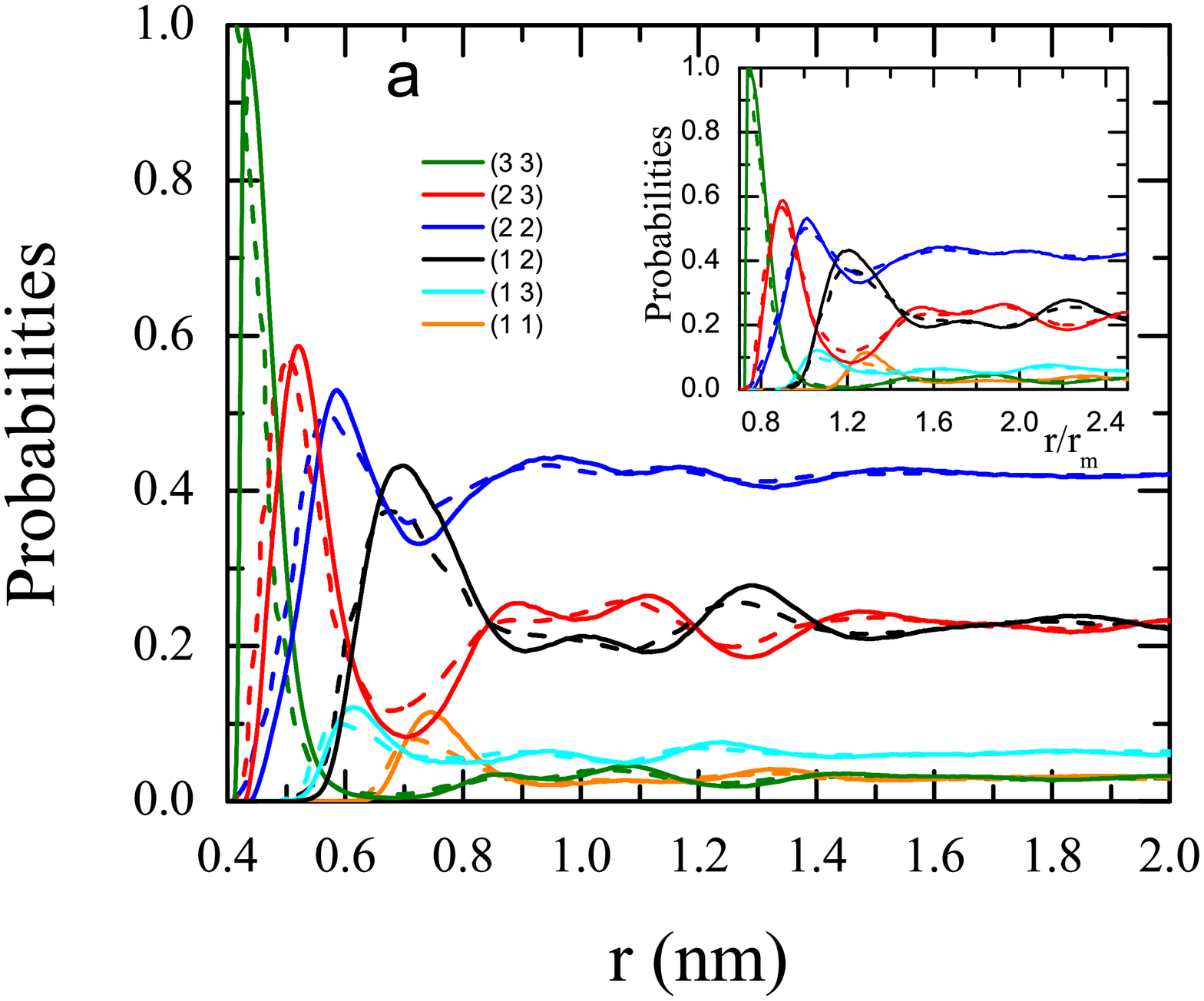}\\
\includegraphics[width=0.4\textwidth]{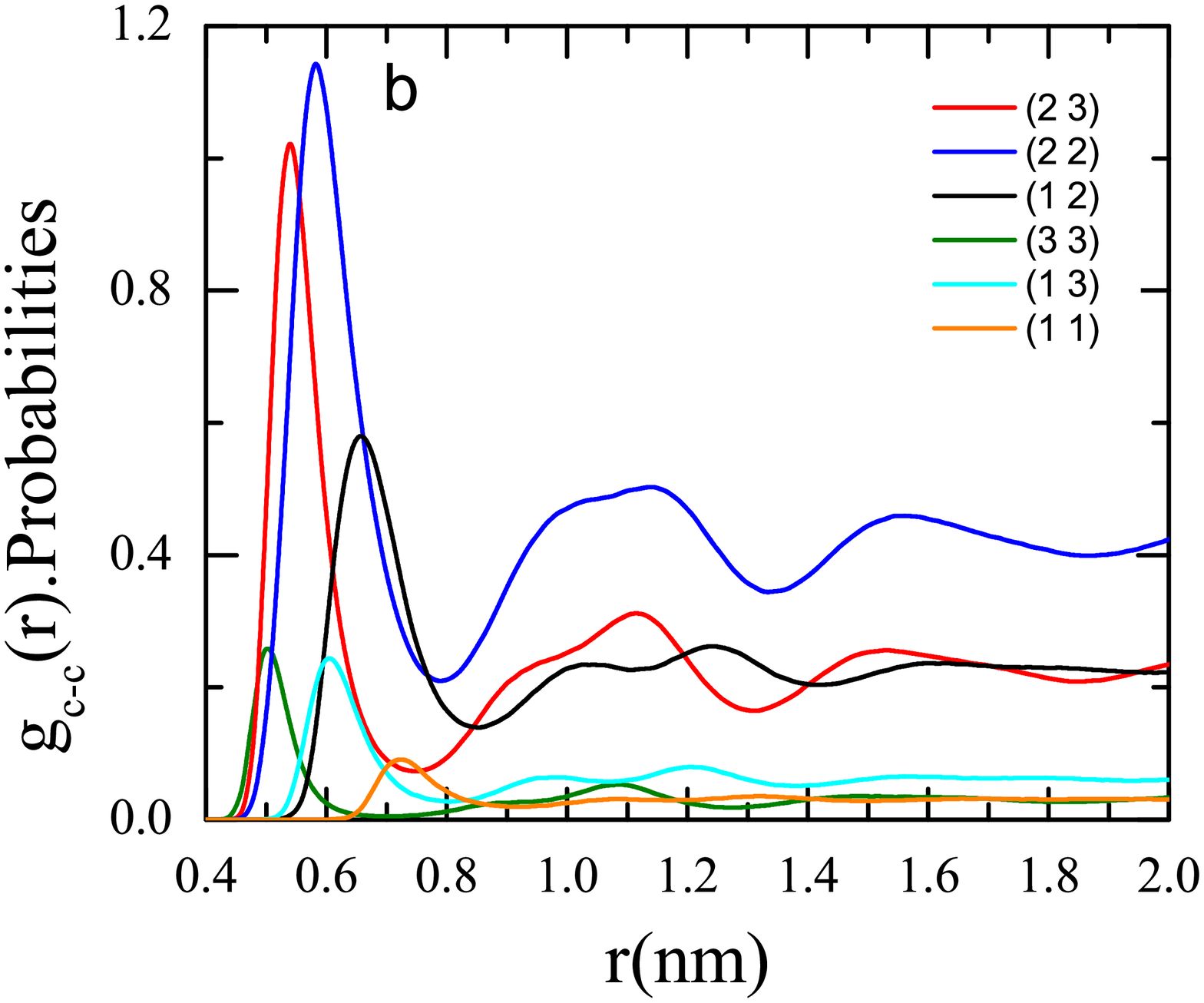}
\end{center}
\caption{  \label{p6BrCl} a) (Color online) Distance dependent probabilities for each of the six classes in CCl$_3$Br (solid lines) and the Rey data for CCl$_4$ (dashed lines). The inset displays an the same probabilities plotted as a function of the reduced distances $r/r_m$ . b) Product of the distance dependent probabilities and the C$-$C radial distribution function for CCl$_3$Br.} 
\end{figure}

We now turn our attention to the results obtained after sorting out all the configurations according to the orientational classification defined above. These results provide details of the structure of the liquid that is absent in the spherically symmetric radial distribution functions. 

In Figure \ref {p6BrCl}a we show the probabilities of each of the six main classes, as a function of the separation between the carbon atoms, corresponding to liquid CCl$_3$Br. For comparison, we also include the findings of Rey for carbon tetrachloride \cite{rey07}. Notice that these probabilities are normalized for each $r$. Even though the symmetry of the two compounds is different ($C_{3v}$ and $T_d$ for trichlorobromomethane and carbon tetrachloride, respectively) they show the same qualitative behavior. The main difference between the two is that the curves for CCl$_3$Br are shifted toward larger $r$ values, reflecting the effect of the larger size of the Br with respect to Cl as mentioned above. Using simple geometrical arguments it easy to visualize that the {\em face-to-face} configuration can be observed for distances at which no other configuration is possible. This is reflected in the probability curves, where this case takes all the probability at closest approach. On the other hand, the {\em corner-to-corner} configuration is only possible for sufficiently large separation, coexisting with all the other configurations. These two cases, plus the combined {\em corner-to-face} configuration, pay the highest entropic price and their asymptotic probabilities are very small. The remaining three configurations display an oscillating pattern spanning first and second neighboring layers, approaching their limiting values for $r \simeq 1$ nm. A more realistic representation of the same results emerges when the probabilities of the different classes is weighted by the radial distribution function, as displayed in Figure \ref{p6BrCl}b. The somewhat misleading high probability shown for by the {\em face-to-face} configuration becomes a small peak in this new representation that indicates the {\em edge-to-edge} conformation as the most abundant, followed by the {\em edge-to-face} and {\em corner-to-edge} at shorter and longer separations, respectively.

The distance dependent relative orientational probabilities for CCl$_3$Br are split into subclasses and shown in Figure 4, for both, liquid and plastic phases. We restrict the analysis to the most abundant configurations; namely, (2,2), (2,3) and (1,2). 
The relative ordering of the different subclasses is the same in the both phases. It is also observed that the proportion of the different subclasses relative to the total of the class is, for all $r$, approximately constant and equal to the asymptotic values.
The main difference is in the position of the different peaks corresponding to the subclasses, which occurs according to the size of the atoms involved. For example, for the (2,2) class, the ordering in $r$ is $(2_0,2_0)$, $(2_1,2_0)$ and $(2_1,2_1)$ for no Br, one Br or two Br between the planes, respectively. In Figure \ref{potbr2} we show probabilities the {\em face-to-face} configuration and the relative order of the subclasses for the liquid and plastic phases. The probability of finding these configurations is very small and only relevant at very short distances, as shown by Figure \ref{p6BrCl}b. However it is interesting that for the liquid the subclass of closest approach is with $k=0$, while for the plastic the three subclasses are all able to reach approximately the same short distance.The positional order in the lattice and its lower thermal energy prevents a definite order of the subclasses. This effect is also reflected by the probability $P_k$ of having confirmations with $k$ substituted atoms between the planes, showed in Figure \ref{pij}a. The overall distance dependent of $P_k$ is similar for the plastic and liquid phases but small differences are observed at very short distances. The liquid clearly allows a closer approach for the configurations with less (or none) Br  atoms between the planes. In the ordered plastic phase, the atoms vibrate about their equilibrium position and the closes approach is independent of $k$. However as it can be seen in Figure \ref{pij}b the probability to find a molecule with $k=0$ is very small at short distances.

\begin{figure}[tt]
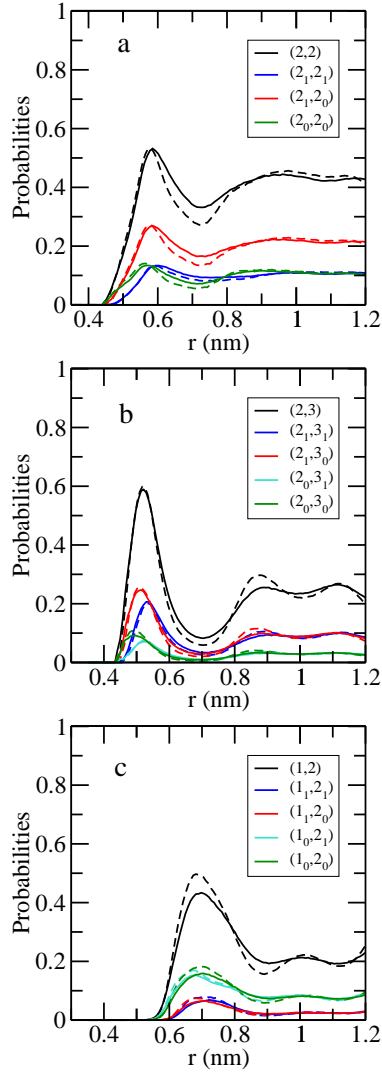

\begin{center}
\includegraphics[width=0.3\textwidth]{sp22.eps}\\
\includegraphics[width=0.3\textwidth]{sp23.eps}\\
\includegraphics[width=0.3\textwidth]{sp12.eps}
\end{center}
\caption{  \label{potbr1} (Color online) Distance dependent probabilities that result for each of the subclasses in a) {\em edge-to-edge}, b) {\em edge-to-face} and c) {\em corner-to-edge} configurations for liquid (solid lines) and plastic (dashed lines) CCl$_3$Br.} 
\end{figure}

\begin{figure}[bt]
\begin{center}
\includegraphics[width=8cm]{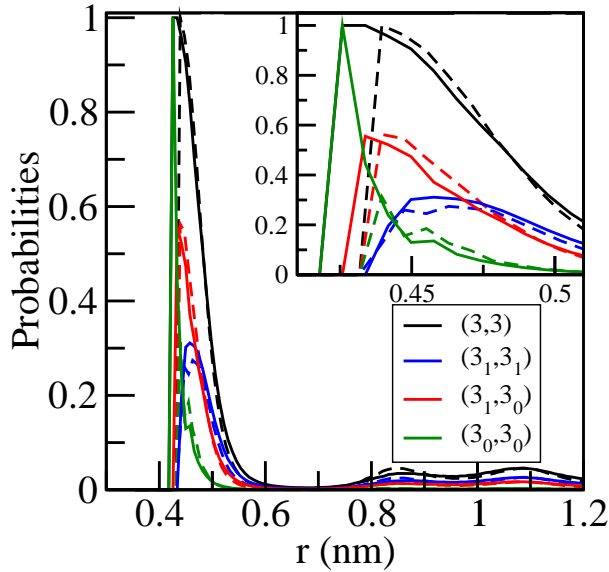}
\end{center}
\caption{  \label{potbr2} (Color online) Distance dependent probabilities that result for each of the subclasses in face to face (3,3) configuration for liquid (solid lines) and plastic (dashed lines) CCl$_3$Br. The inset shows details at short distances.} 
\end{figure}

\begin{figure}[bt]
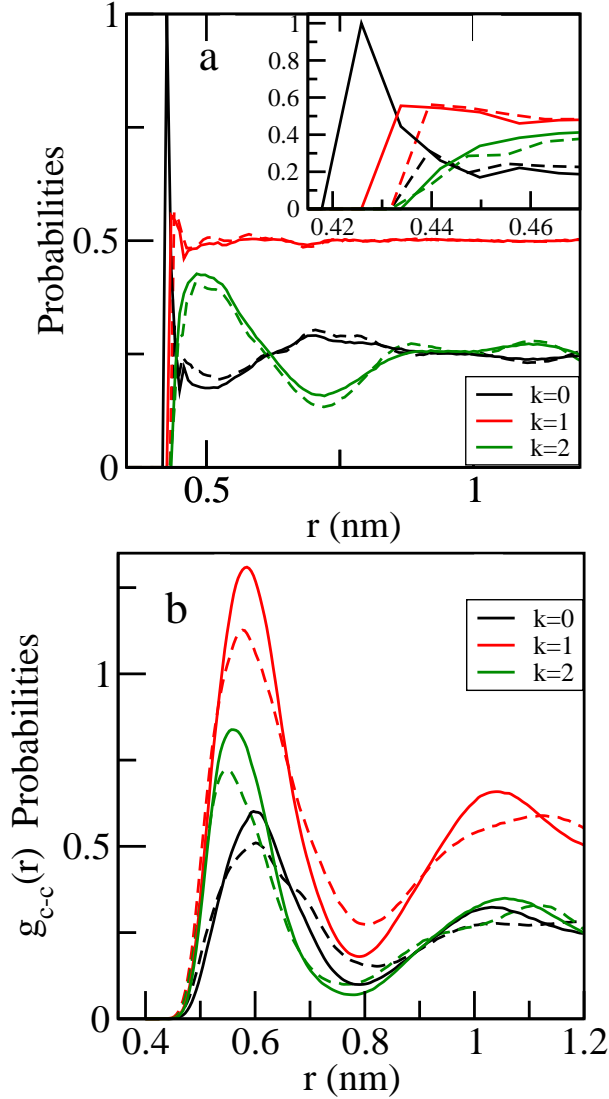

\begin{center}
\includegraphics[width=7.3cm]{Fig6af.eps} \\
\includegraphics[width=8cm]{Fig6bf.eps}
\end{center}
\caption{  \label{pij} (Color online) a) Distance dependent probabilities $P_k$ of having $k=0,1,2$ substituted atoms between planes for liquid (solid lines) and plastic (dashed lines) CCl$_3$Br. The inset shows details at short distances. b) Product of the probabilities given in (a) times the C-C radial distribution function.} 
\end{figure}

\begin{figure}[bt]
\begin{center}
\includegraphics[width=12cm]{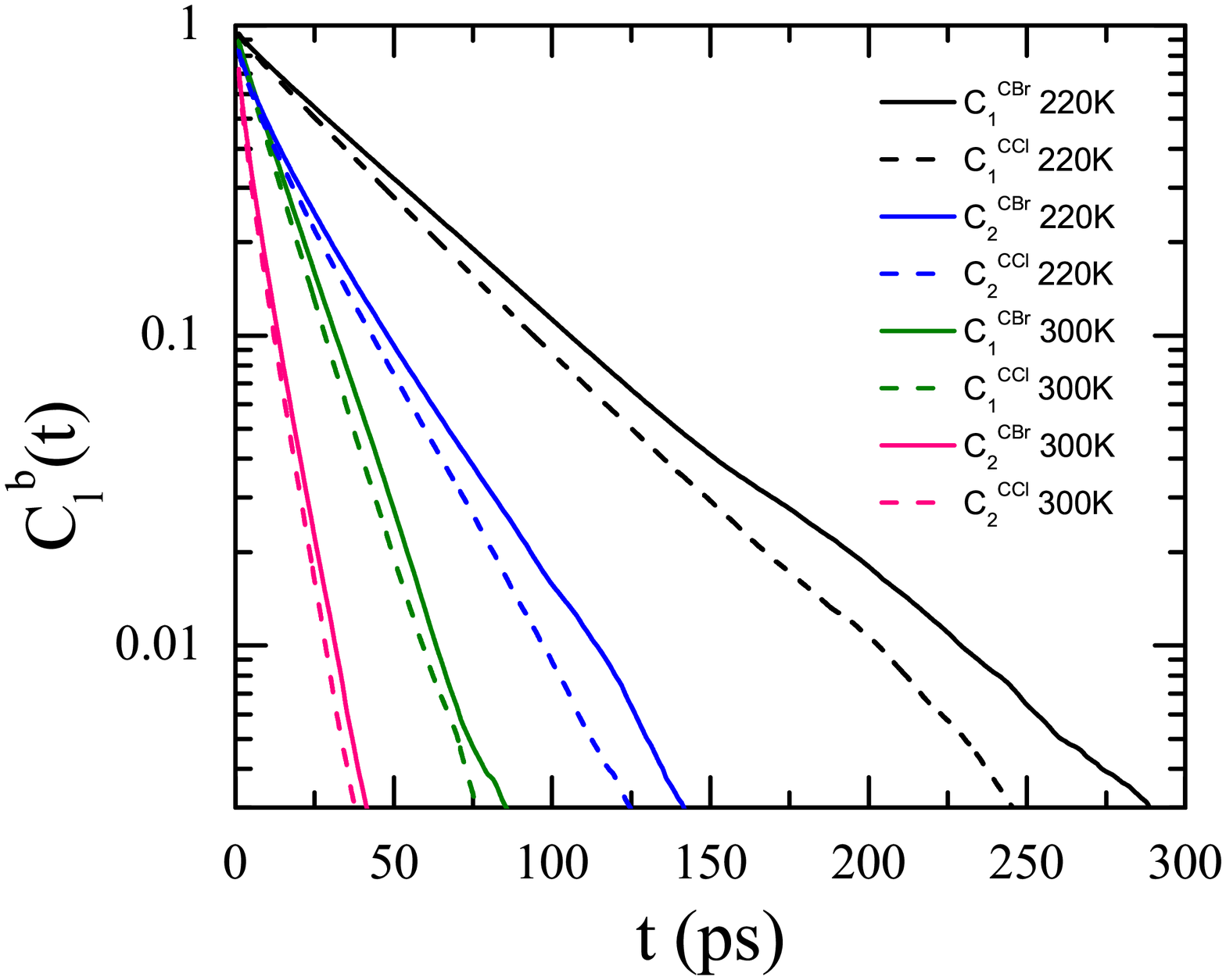}
\end{center}
\caption{  \label{correl} (Color online) Correlation functions $C_l^b(t)$ for $l= 1, 2$ and $b= Cl, Br$ for the liquid phase (300 K) and the plastic phase (T=220K)}.
\end{figure}


\subsection{Rotational relaxation. Single molecule dynamics}

The rotational dynamics of molecular systems can be investigated using appropriate time correlations functions. For the case of CCl$_3$Br we used correlations between selected bonds in each molecule. Let us define $\vec{u}^{CBr}$ a unit vector oriented along the C$-$Br bond and r $\vec{u}^{CCl}$ a unit vector oriented along one of the C$-$Cl bond. Then we define the correlation function
\begin{equation}
C_l^b(t)\,=\, \frac{1}{N} \sum_i
\left\langle P_l\left(\vec{u}_i^b(0) \cdot \vec{u}_i^b(t) \right) 
\right\rangle  \,\,\, ,
\end{equation}
where $P_l$ is the $l$-order Legendre polynomial, $i$ runs over all the molecules, and $\vec{u}_i^b$  is a unit vector directed along the C$-b$  bond ($b=$ Cl,Br) of the molecule $i$. The analysis of these correlation functions shed information about the relative rotational preferences of the molecules, and the time scale of these rotations. 

\begin{figure}[bt]
\begin{center}
\includegraphics[width=10cm]{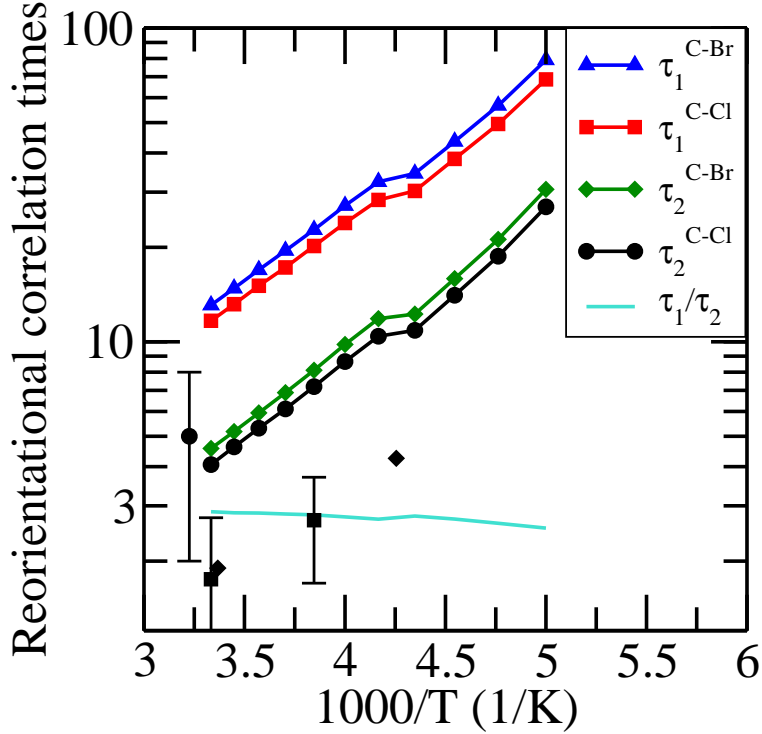}
\end{center}
\caption{  \label{taus} (Color online) CCl$_3$Br rotational correlation times $\tau_1$ and $\tau_2$ in liquid and plastic phases deduced from the reorientational correlation functions for $\vec{u}_{CCl}$ (red and black) and $\vec{u}_{CBr}$ (blue and green). Cyan lines correspond to the ratios $\tau_1^b$/$\tau_2^b$. Square and diamonds experimental \cite{gillen72b} and MD \cite{rey08b} values of $\tau_2$ for CCl$_4$,circle experimental value of $\tau_2$ for CBr$_4$ \cite{more84} .} 
\end{figure}

In Figure \ref{correl} we show the correlation functions $C_l^b(t)$ for the liquid and plastic phases of CCl$_3$Br and $l=1, 2$. All correlations decay monotonically over a time scale of 10 to 100 ps with an approximately exponential behavior. The liquid systems show a faster decay than the plastic phase, as expected. The decay is very similar along both types of bonds, indicating a similar dynamics around the different bonds. The rotational correlation time $\tau_l$ for the correlation function $C_l^b(t)$  can be extracted by different methods: $i$) exponential fit of the correlation functions, $ii$) integration of the correlation function or $iii$) time at which the correlation function decay to 1/$e$. All three methods yield practically indistinguishable results. In Figure \ref{taus} we show the rotational correlation times as a function of temperature. For comparison, we also include experimental values of $\tau_2$ for CCl$_4$ \cite{gillen72b,rey08b} and CBr$_4$ \cite{more84} in the liquid and plastic phase, since there are no experiments for CCl$_3$Br reported in the literature.

\begin{figure}[bt]
\begin{center}
\includegraphics[width=12cm]{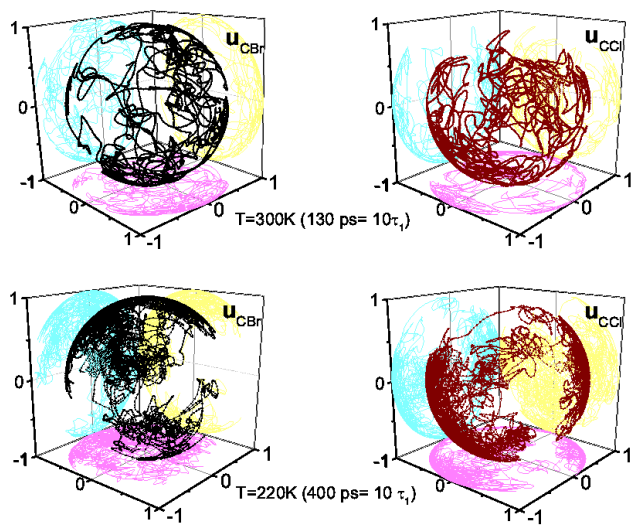}
\end{center}
\caption{  \label{esfs} (Color online) Trajectories  of the unitary vectors $\hat{u}_{CBr}$ (black) and $\hat{u}_{CCl}$ (red) of one arbitrary molecule  at two temperatures T = 300 K (liquid phase) and T = 220 K (plastic phase)for 10 $\tau_1$ ps of the simulation run. The rotational motion looks isotropic only at high temperature.} 
\end{figure}

In the isotropic rotational diffusion model it is assumed that the rotation of the molecule occurs trough random discrete jumps in all possible directions with equal probability. Then, after a waiting time longer that the correlation time, the molecules show no preferred orientation. It has been demonstrated that for this model $\tau_1/\tau_2=3$ \cite{debye29}. As it can be seen in Figure \ref{taus} this ratio decreases from 2.9 at high temperature (liquid) to 2.5 for the lowest temperature system (plastic). This indicates that the rotational motion is nearly isotropic in the liquid phase and some departure from this behavior is suggested in the plastic phase.
In order to verify weather the rotational motion is truly isotropic or not, we examine the trajectory of the bonds vectors $\vec{u}_{CBr}$ and $\vec{u}_{CCl}$ of one of the molecules. These trajectories and their projections over the  three coordinate planes are shown in Figure \ref {esfs}. It is clear that in the liquid the path traced by the tips of both vectors defines the surface of a sphere showing that the molecule visit nearly all the possible orientations during a time longer than the rotational correlation time. However this is not the case in the low temperature plastic phase were it is observed that only some regions of the sphere are covered, indicating that in this case the motion is not truly isotropic.

\section{Conclusions}

We have studied by molecular dynamics techniques the short range order and the rotational relaxation dynamics of CCl$_3$Br in the liquid and plastic phase.

The relative orientational order of the CCl$_3$Br molecules was investigated using the classification defined by Rey \cite{rey07} for tetrahedral molecules and extended by Pothoczki et al \cite{pothoczki11} for molecules with $C_{3v}$ symmetry. 
The findings from our MD simulations are in good agreement with reverse Monte Carlo results in $C_{3v}$ molecules  reported in reference  \cite{pothoczki11}, providing a validation for the our molecular models.  As expected, since CCl$_3$Br is a slightly distorted tetrahedron, our results for the main six classes are in qualitative agreement with those for CCl$_4$ that is a perfect tetrahedron. However, we have shown that even though the configuration {\em face-to-face} has a maximum  at very short distances, the probability to find two molecules at those distances is very small. A similar phenomenon occurs with the probabilities $P_k$ to find $k= 0,1,2$ substituted atoms between planes. The difference between liquid and plastic phase is mainly observed at those short distances, less than 0.45 nm, for which $g_{c-c}(r)$ is almost zero.

From the correlation functions of the C-Br and C-Cl bond vectors we have found the rotational relaxation times $\tau_1$ and $\tau_2$. The values of $\tau_2$  determined from our MD simulations are in the range of values found in CCl$_4$ and CBr$_4$. The temperature dependence is consistent with an Arrhenius behavior in liquid and plastic phases and very similar to that of CCl$_4$. The ratio $\tau_1/\tau_2$  indicates than the rotational motion is isotropic and diffusive in the liquid phase but non isotropic in the plastic phase as it was shown to happen in another tetrahedral molecules such as neopentane \cite{rey08a} and CBr$_4$ \cite{descamps84}.

\acknowledgments NBC, MZ and PS  would like to acknowledge SECYT-UNC, CONICET,
and MINCyT C\'ordoba for partial financial support of this project.


\begin{thebibliography}{99}

\bibitem{parsonage78} N.G. Parsonage and L.A.K. Staveley, {\em Disorder in Crystals},
Clarendon Press, Oxford (1978).

\bibitem{sherwood79} J.N. Sherwood, {\em The Plastically Crystalline State: Orientationally Disordered Crystals}, Wiley, New York (1979).

\bibitem{gillen72} K.T. Gillen, J.E. Griffiths,
 Chem. Phys. Lett.  {\bf 17}, 359 (1972).

\bibitem{anderson79} A. Anderson, B.H. Torrie and W.S. Tse,
Chem. Phys. Lett. {\bf 61}, 119 (1979).

\bibitem{cohen79} S. Cohen, R. Powers and R. Rudman, Acta Crystallogr.
{\bf B 35}, 1670 (1979).

\bibitem{temleitner10} L. Temleitner and L. Pusztai, Phys. Rev. B {\bf 81}, 134101 (2010).

\bibitem{llanta01} E. Llanta and R. Rey, Chem. Phys. Lett. {\bf 340}, 173 (2001).

\bibitem{pardoPRB2007} L. C. Pardo, J. Ll. Tamarit, N. Veglio, F. J. Bermejo and G. J. Cuello, Phys. Rev. B {\bf 76}, 134203 (2007).

\bibitem{reyjcp2009} R. Rey, J. Chem. Phys. {\bf 131}, 064502 (2009).

\bibitem{rey08a} R. Rey, J.  Phys. Chem. B {\bf 112}, 344 (2008).

\bibitem{rey08b} R. Rey, J. Chem. Phys. {\bf 129}, 224509 (2008).

\bibitem{zuriaga09} M. Zuriaga, L. C. Pardo, P. Lunkenheimer, J. L. Tamarit,
N. Veglio, M. Barrio, F. J. Bermejo y A. Loidl,
Phys. Rev. Lett. {\bf 103}, 075701 (2009).

\bibitem{zuriaga11} M. Zuriaga, M. Carignano and P. Serra, J. Chem. phys. {\bf 135}, 044504 (2011).

\bibitem{mcdonald82} I.R. McDonald, D.G. Bounds and M.L. Klein, Mol. Phys. {\bf 45}, 521 (1982).

\bibitem{dove86} M.T. Dove, J. Phys. C {\bf 19}, 3325 (1986).

\bibitem{tironi96} I. G. Tironi, P. Fontana and W.F. Van Gunsteren, Mol. Sim.
{\bf 18}, 1 (1996).

\bibitem{pettitt79} B.A. Pettitt and R.E. Wasylishen, Chem. Phys. Lett.
{\bf 63}, 539 (1979).

\bibitem{ohta95} T. Ohta, O. Yamamuro and T. Matsuo, J. Phys. chem. {\bf 99},
2403 (1995).

\bibitem{binbrek99} O.S. Binbrek, S.E. Lee-Dadswell, B.H. Torrie and B.M. Powell, Mol. Phys. {\bf 96}, 785 (1999).

\bibitem{parat05} B. Parat, L.C. Pardo, M. Barrio, J.L. Tamarit, P. Negrier,
J. Salud, D.O. L\'opez and D. Mondieig, Chem. Mater. {\bf 17}, 3359 (2005).

\bibitem{pothoczki10} S. Pothoczki, L. Temleitner and L. Pusztai,
J. Chem. Phys. {\bf 132}, 164511 (2010).

\bibitem{pothoczki11}  S. Pothoczki, L. Temleitner and L. Pusztai,
J. Chem. Phys. {\bf 134}, 044521 (2011).


\bibitem{rey07} R. Rey, J. Chem. Phys. {\bf 126}, 164506 (2007).



\bibitem{rey00} R. Rey, L.C. Pardo, E. Llanta, K. Ando, D.O. L\'opez,
J.L. Tamarit and M. Barrio, J. Chem. Phys. {\bf 112}, 7505 (2000).

\bibitem{Yama98} O. Yamamuro, T. Ohtay and T. Matsuo, J. Korean Phys. Soc. {\bf 32}, S838 (1998).

\bibitem{Miller56} R. C. Miller and C. P. Smyth, J. Chem. Phys. {\bf 24}, 814 (1956).


\bibitem{evans85} M. W. Evans and G. J. Evans, Adv. Chem. Phys {\bf 63}, 377 (1985).

\bibitem{barclay92} B. J. Barclay, D. B. Jack, J. C. Polanyi and Y. Zeiri, J. Chem. Phys., {\bf 97}, 9458 (1992).

%
\bibitem{gromacs}
D. van~der Spoel, E.  Lindahl, B. Hess, G. Groenhof, A. Mark, \& H. Berendsen,
{\bf 26},   1701 (2005).
%
\bibitem{gromacs4}  B. Hess, C. Kutzner, D. van~der Spoel  \& E. Lindahl,
JCTC {\bf 4}, 435 (2008).
%
\bibitem{gillen72b} K. Gillen, J. H. Noggle and T. K. Leipert, Chem. Phys. Lett. {\bf 17}, 505 (1972).

\bibitem{more84} M. More, J. Lefebvre and B. Hennion, J. Physique {\bf 45}, 303 (1984).

\bibitem{debye29} P. Debye, Polar Molecules (Dover, New York, 1929).

\bibitem{descamps84} M. Descamps, J. Physique {\bf 45}, 587 (1984).


\end{thebibliography}
\end{document}